\documentclass[aps,twocolumn,groupedaddress,showpacs,prb,floatfix]{revtex4}
\usepackage{graphicx,rotating,subfigure,amsmath,amsfonts,amssymb,delarray}

\newcommand{\e}{\text{e}}
\newcommand{\im}{\text{i}}

\begin{document}
\bibliographystyle{apsrev}


\title{Real-time dynamics at finite temperature by DMRG:\\
A path-integral approach}


\author{Jesko Sirker}
\affiliation{Department of Physics and Astronomy, University of British
  Columbia, Vancouver, British Columbia, Canada V6T 1Z1}
\author{Andreas Kl\"umper}
\affiliation{Theoretische Physik, Universit\"at Wuppertal, Gau{\ss}-Str. 20,
  42097 Wuppertal, Germany}


\date{\today}

\begin{abstract}
%
%
%
  We propose a path-integral variant of the DMRG method to calculate real-time
  correlation functions at arbitrary finite temperatures. To illustrate the
  method we study the longitudinal autocorrelation function of the
  $XXZ$-chain. By comparison with exact results at the free fermion point we
  show that our method yields accurate results up to a limiting time which is
  determined by the spectrum of the reduced density matrix.
\end{abstract}

\maketitle

The density-matrix renormalization group (DMRG) \cite{WhiteDMRG} is today a
well established numerical method to study ground state properties of
one-dimensional quantum systems. Within the last few years the DMRG method has
been generalized to allow also for the calculation of spectral functions
\cite{JeckelmannDDMRG} and, quite recently, to incorporate directly real-time
evolution.\cite{WhiteFeiguin} The most powerful variant of DMRG to calculate
thermodynamic properties is the density-matrix renormalization group applied
to transfer matrices (TMRG). This method has been proposed by Bursill {\it et
  al.} \cite{BursillXiang} and has later been improved
considerably.\cite{WangXiang} The main idea of TMRG is to express the
partition function $Z$ of a one-dimensional quantum model by that of an
equivalent two-dimensional classical model obtained by a Trotter-Suzuki
decomposition.\cite{Trotter} Thermodynamic quantities can then be calculated
by considering a suitable transfer matrix $\mathcal{T}$ for the classical
model.  The main advantage of this method is that the thermodynamic limit
(chain length $L\rightarrow\infty$) can be performed exactly and that the free
energy in the thermodynamic limit is determined solely by the largest
eigenvalue of $\mathcal{T}$. The TMRG has been applied to calculate static
thermodynamic properties for a variety of one-dimensional systems including
spin chains, the Kondo lattice model, the $t-J$ chain and ladder, and also
spin-orbital models.\cite{EggertRommer,MutouShibata,SirkerKluemperPRB}

The Trotter-Suzuki decomposition of a one-dimensional quantum system yields a
two-dimensional classical model with one axis corresponding to imaginary time
(inverse temperature). It is therefore straightforward to calculate
imaginary-time correlation functions (CFs) using the TMRG algorithm. Although
the results for the imaginary-time CFs obtained by TMRG are very accurate, the
results for real-times (real-frequencies) involve large errors because the
analytical continuation poses an ill-conditioned problem. In practice it has
turned out that the maximum entropy method is the most efficient and reliable
way to obtain spectral functions from TMRG data. The combination of TMRG and
maximum entropy has been used to calculate spectral functions for the
$XXZ$-chain \cite{NaefWang} and the Kondo-lattice model.\cite{MutouShibata}
However, this method involves intrinsic errors due to the analytical
continuation which cannot be resolved.

Here we propose a method to calculate directly real-time CFs at finite
temperature by a modified TMRG algorithm thus avoiding an analytical
continuation. We start by considering the two-point CF for an operator
$\hat{O}_r(t)$ at site $r$ and time $t$
\begin{eqnarray}
\label{eq1}
\langle\hat{O}_r(t)\hat{O}_0(0)\rangle &=&
\frac{\mbox{Tr}\left(\hat{O}_r(t)\hat{O}_0(0)\e^{-\beta
      H}\right)}{\mbox{Tr}\left(\e^{-\beta H}\right)} \\
&=& \frac{\mbox{Tr}\left(\e^{-\beta
      H/2}\e^{\im t H}\hat{O}_r\e^{-\im t H}\hat{O}_0\e^{-\beta
      H/2}\right)}{\mbox{Tr}\left(\e^{\im t H}\e^{-\im tH}\e^{-\beta
      H}\right)} \;  \nonumber
\end{eqnarray} 
where $\beta$ is the inverse temperature. Here we have used the cyclic
invariance of the trace and 
%
%
%
have written the denominator in analogy to the numerator.
For a Hamiltonian $H$ with nearest-neighbor interactions we can use the
Trotter-Suzuki decomposition
\begin{equation}
\label{eq2}
\e^{-\epsilon H} = \e^{-\epsilon H_{odd}/2} \e^{-\epsilon H_{even}} \e^{-\epsilon H_{odd}/2} +\mathcal{O}(\epsilon^3)
\end{equation} 
where $\epsilon = \beta/M$ so that the partition function $Z=\exp(-\beta H)$ becomes
\begin{equation}
\label{eq3}
Z = \text{Tr} \left( \prod_{i= \text{odd}}\e^{-\epsilon
    h_{i,i+1}}\prod_{i=\text{even}}\e^{-\epsilon h_{i,i+1}}\right)^M
\left(1+\mathcal{O}(\epsilon^2)\right) \; .
\end{equation}
With $\im t \rightarrow \tau$ in Eq.~(\ref{eq1}) and inserting the identity
operator at each imaginary time step one obtains directly a lattice
path-integral representation for the imaginary time CF
$\langle\hat{O}_r(\tau)\hat{O}_0(0)\rangle$.\cite{NaefWang,MutouShibata}

\begin{figure}[!htp]
\begin{center}
\includegraphics*[width=0.99\columnwidth]{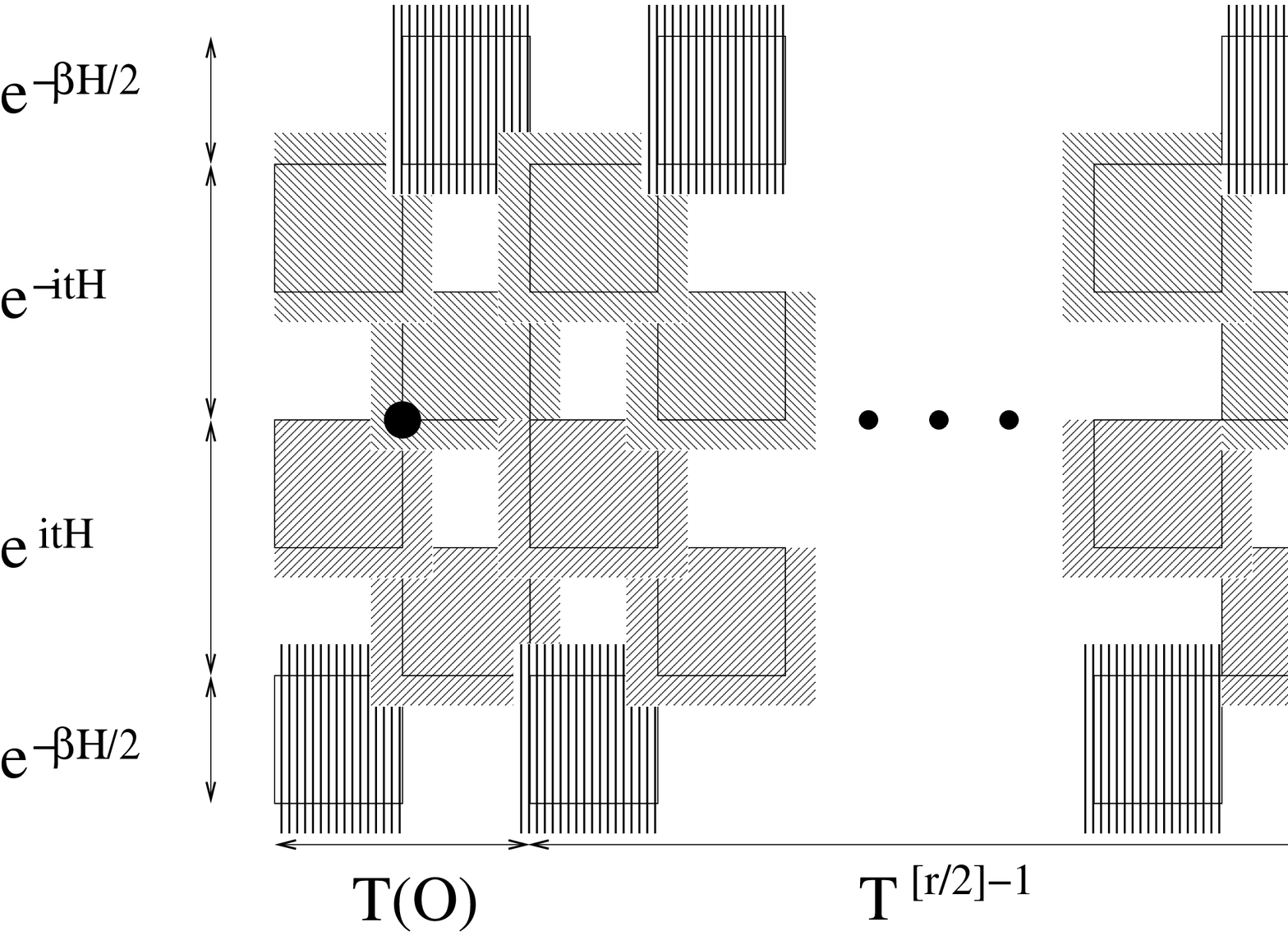}
\end{center}
\caption{Transfer matrices appearing in the numerator of Eq.~(\ref{eq6}) for $r>1$
  with $r$ even. The 2 black dots denote the operator $\mathcal{O}$. Each
  transfer matrix consists of three parts: A part representing $\exp(-\beta
  H)$ (vertically striped plaquettes), another for $\exp(\im t H)$ (stripes
  from lower left to upper right) and a third part describing $\exp(-\im t H)$
  (upper left to lower right). The transfer matrices
  $\mathcal{T},\mathcal{T}(O)$ are split into a system (S) and an environment
  block (E).}
\label{fig1}
\end{figure}

The crucial step in our new approach for real times is to introduce a second
Trotter-Suzuki decomposition of $\exp(-\im\delta H)$ as in Eq.~(\ref{eq2})
with $\delta=t/N$. We can then define a column-to-column transfer matrix 
\begin{eqnarray}
\label{eq4}
\mathcal{T}_{2N,M} &=&
(\tau_{1,2}\tau_{3,4}\cdots\tau_{2M-1,2M})(\tau_{2,3}\tau_{4,5}\cdots\tau_{2M,2M+1})
\nonumber \\
&&
(\bar{v}_{2M+1,2M+2}\cdots\bar{v}_{2M+2N-1,2M+2N}) \nonumber \\
&& (\bar{v}_{2M+2,2M+3}\cdots\bar{v}_{2M+2N,2M+2N+1})
\nonumber \\
&& (v_{2M+2N+1,2M+2N+2}\cdots v_{2M+4N-1,2M+4N}) \nonumber \\
&& (v_{2M+2N+2,2M+2N+3}\cdots v_{2M+4N,1})
\end{eqnarray}
%
%
where the local transfer matrices have matrix elements
\begin{eqnarray}
\label{eq5}
\tau(s_k^i s_k^{i+1}| s_{k+1}^i s_{k+1}^{i+1}) &=& \langle s_k^i s_k^{i+1} | \e^{-\epsilon h_{i,i+1}} |
  s_{k+1}^i s_{k+1}^{i+1} \rangle \\
v(s_l^i s_l^{i+1}| s_{l+1}^i s_{l+1}^{i+1}) &=& \langle s_l^i s_l^{i+1} | \e^{-\im\delta h_{i,i+1}} |
  s_{l+1}^i s_{l+1}^{i+1} \rangle \;  \nonumber 
\end{eqnarray}
and $\bar{v}$ is the complex conjugate. Here $i=1,\cdots,L$ is the lattice
site, $k=1,\cdots,2M$ ($l=1,\cdots,2N$) 
%
%
the index of the imaginary time
(real time) slices and $s_{k(l)}^i$ denotes a local basis. The denominator in
Eq.~(\ref{eq1}) can then be represented by $\mbox{Tr}(\mathcal{T}_{2N,M}^{L/2})$ where
$N,M,L\rightarrow\infty$. A similar path-integral representation 
%
%
holds for the numerator in (\ref{eq1}). 
%
%
Here we have to introduce an additional modified transfer matrix
$\mathcal{T}_{2N,M}(\hat{O})$ which contains the operator $\hat{O}$ at the
appropriate position. For $r>1$ we find
\begin{eqnarray}
\label{eq6}
&&\langle\hat{O}_r(t)\hat{O}_0(0)\rangle  \nonumber \\
&&=\lim_{N,M\rightarrow\infty}\lim_{L\rightarrow\infty}\frac{\mbox{Tr}(\mathcal{T}(\hat{O})\mathcal{T}^{[r/2]-1}\mathcal{T}(\hat{O})\mathcal{T}^{L/2-[r/2]-1})}{\mbox{Tr}(\mathcal{T}^{L/2})}
 \nonumber \\
&&= \lim_{N,M\rightarrow\infty}
\frac{\langle\Psi^L_0|\mathcal{T}(\hat{O})\mathcal{T}^{[r/2]-1}\mathcal{T}(\hat{O})|\Psi^R_0\rangle}{\Lambda_0^{[r/2]+1}\langle\Psi^L_0|\Psi^R_0\rangle}
 \; . 
\end{eqnarray}   
%
%
Here $[r/2]$ denotes the first integer smaller than or equal to $r/2$ and we
have set $\mathcal{T}\equiv \mathcal{T}_{2N,M}$. In the second step we have
used the fact that the spectrum of the column-to-column transfer matrix
$\mathcal{T}$ is gapped for $\beta < \infty$.\cite{WangXiang} Therefore the
trace is reduced to an expectation value where $|\Psi^R_0\rangle,
\langle\Psi^L_0|$ are the right- and left-eigenvectors of the non-hermitian
transfer matrix $\mathcal{T}$ belonging to the largest eigenvalue
$\Lambda_0$. A graphical representation of the transfer matrices appearing in
the numerator of Eq.~(\ref{eq6}) is shown in Fig.~\ref{fig1}.

It is worth to note that Eq.~(\ref{eq6}) has the same structure as the
equation for the calculation of imaginary time
CFs.\cite{NaefWang,MutouShibata} Only the transfer matrices involved are
different. For practical DMRG calculations the parameters $\epsilon,\delta$
are fixed and the temperature (time) is decreased (increased) by increasing M
(N). This is achieved by splitting $\mathcal{T},\mathcal{T}(O)$ into a system
and an environment block (see Fig.~\ref{fig1}). Adding an additional
$\tau$-plaquette at the lower end of the system block then decreases the
temperature whereas adding a $v$-plaquette at the upper end leads to an
increase in time $t$. The extended system block is renormalized by projecting
onto the $Z$ largest eigenstates of the reduced density matrix $\rho_S =
\mbox{Tr}_E |\Psi^R_0\rangle\langle\Psi^L_0|$.  Starting from an existing TMRG
program our real-time algorithm requires only the following minor changes: In
addition to the $\tau$-plaquettes also $v$-plaquettes are present. Therefore
the transfer matrix and the density matrix become complex quantities so that
complex diagonalization routines are required.
%
%
%

In the remainder of this letter we want to study, as a highly non-trivial
example, the longitudinal autocorrelation $C(t,T)=\langle
S^z_0(t)S^z_0(0)\rangle$ at temperature $T$ for the $XXZ$ chain with
Hamiltonian
%
%
\begin{equation}
\label{eq7}
H = J\sum_i \left(S^x_iS^x_{i+1}+S^y_iS^y_{i+1}+\Delta S^z_iS^z_{i+1}\right)
\end{equation} 
where $S=1/2$, $J>0$ and $\Delta\geq 0$. Although the model is integrable,
exact results for $C(t,T)$ are available only at the free fermion point
$\Delta = 0$. Quite interestingly, $C(t,T)$ shows a non-trivial behavior even
at infinite temperature due to the quantum nature of the
problem.\cite{FabriciusMcCoy}

\begin{figure}[!htp]
\begin{center}
\includegraphics*[width=0.99\columnwidth]{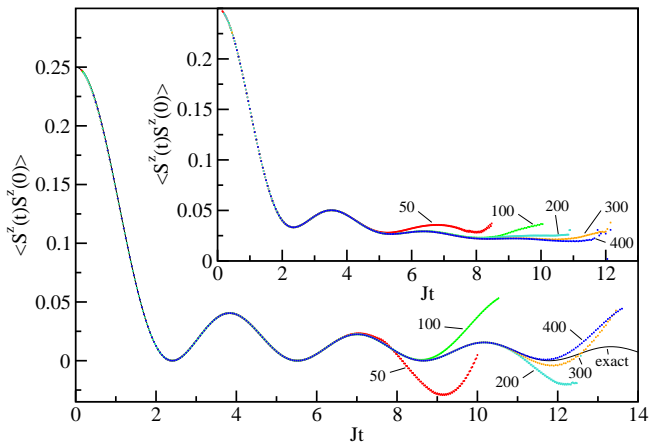}
\end{center}
\caption{Longitudinal autocorrelation for $\Delta=0$ and $\Delta=1$ (inset) at
  $T=\infty$ where the number of states kept varies between $Z=50$ and $400$
  and $\delta=0.1$. The exact result is shown for comparison in the case
  $\Delta=0$.}
\label{fig2}
\end{figure}

For the autocorrelation both $S^z$ operators are situated in the same transfer
matrix $\mathcal{T}(S^z,S^z)$ so that Eq.~(\ref{eq6}) reduces to
\begin{equation}
\label{eq8}
C(t,T)=
\frac{\langle\Psi^L_0|\mathcal{T}(S^z,S^z)|\Psi^R_0\rangle}{\Lambda_0\langle\Psi^L_0|\Psi^R_0\rangle}
\; .
\end{equation}
It is important to note that once the blocks necessary to construct
$\mathcal{T}, \mathcal{T}(S^z,S^z)$ at a given time $t$ are known, the
correlation function for arbitrary distance $r$ with $t$ fixed can be simply
calculated by additional matrix-vector multiplications (see Eq.~(\ref{eq6})).
We start with the case $T=\infty$ where $\mathcal{T},\mathcal{T}(S^z,S^z)$ do
not contain any $\tau$-plaquettes. This limit directly addresses the essential
feature in our approach, namely the transfer-matrix representation of the time
evolution operator. In Fig.~\ref{fig2} results for $\Delta=0$ and $\Delta=1$
are shown where the number of states kept in the DMRG varies between $Z=50$
and $Z=400$.
In the case $\Delta=0$ the autocorrelation function can be calculated exactly
by mapping the system to a free spinless fermion model using the Jordan-Wigner
transformation. For arbitrary temperature $T$ the result can be expressed as
$C(t,T) = \{D_1(t)+D_2(t,T)\}^2$ with
\begin{equation}
\label{eq9}
D_1(t)=\frac{J_0(Jt)}{2}\; ,\; 
D_2(t,T) = \frac{\im}{\pi}\int_0^1\frac{\sin J\theta t}{\sqrt{1-\theta^2}}\tanh\frac{\theta}{2T}d\theta
\end{equation}
where $J_0$ is the Bessel function of order zero.\cite{Niemeijer} At
$T=\infty$ this reduces to $C(t,\infty) = J_0^2(Jt)/4$. Our numerical results
agree with the exact one up to a maximum time which is determined by the
number of states kept in the DMRG. For the maximum number of states, $Z=400$,
considered here we are able to reproduce the exact result up to $Jt\sim 12$.
For $\Delta=1$ no exact result is available to compare with, however, the
$\Delta=0$ case suggests that the results are trustworthy at least as long as
they agree with the data where a much smaller number of states is kept. The
numerical data with $Z=400$ should therefore be correct at least up to $Jt\sim
10$. We also checked the $\Delta=1$ data for small $t$ by comparing with exact
diagonalization. For TMRG calculations the free fermion point is in no way
special and the algorithm is expected to show the same behavior at this point
as for general $\Delta$.\cite{SirkerKluemperPRB} In the following we will
therefore concentrate on $\Delta=0$ where a direct comparison with exact
results is possible.

\begin{figure}[!htp]
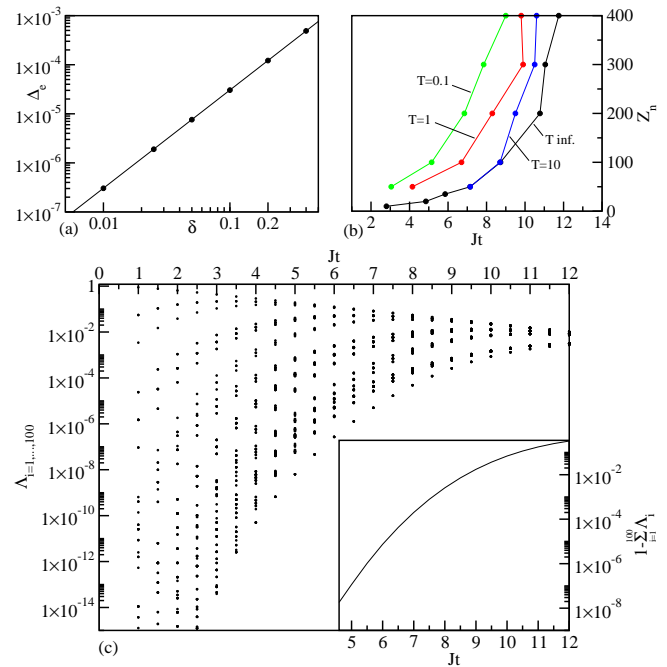

\begin{center}
\includegraphics*[width=0.99\columnwidth]{Fig3ab.eps}
\includegraphics*[width=0.99\columnwidth]{Fig3c.eps}
\end{center}
\caption{(a) $\Delta_e=|C(t,\infty)_{exact}-C(t,\infty)_{TMRG}|$ at $t=4$ as a
  function of $\delta$ with $Z=100$ states kept. The line corresponds to a fit
  with $\Delta_e= 3.05\cdot 10^{-3}\delta^2$. (b) Number of states $Z_n$
  required to keep the error below $10^{-3}$ as function of $t$ for
  $T/J=\infty,10,1,0.1$. (c) Largest 100 eigenvalues $\Lambda_i$ of $\rho_S$
  at $T=\infty$ calculated exactly. The inset shows the discarded weight
  $1-\sum_{i=1}^{100} \Lambda_i$.}
\label{fig3}
\end{figure}

For small $t$ the error in the numerics is entirely dominated by the finite
Trotter-Suzuki decomposition parameter $\delta$. It is therefore possible to
enhance the accuracy by decreasing $\delta$ as shown in Fig.~\ref{fig3}(a). As
expected, the error is quadratic in $\delta$ but fortunately with a rather
small pre-factor $\sim 10^{-3}$. For small $\delta$ more RG steps are
necessary to reach the same $t$. Interestingly, the breakdown of the algorithm
does not depend on the number of RG steps. For different $\delta$ but fixed
$Z$ it always occurs at about the same time $t_c$. Furthermore the breakdown
is always a very rapid one, i.e., for times considerably larger than $t_c$ the
errors become arbitrarily large. This suggests that there is an intrinsic
maximum time scale set by the problem itself. This is supported by
Fig.~\ref{fig3}(b) showing a rapid increase in the number of states $Z_n$
necessary to keep the error below $10^{-3}$.
To understand this behavior we have calculated the spectrum of the reduced
density matrix $\rho_S$ exactly for $T=\infty$ and $\Delta=0$. The result is
shown in Fig.~\ref{fig3}(c). For small $t$ the spectrum is decaying rapidly so
that indeed a few states are sufficient to represent the transfer matrix
$\mathcal{T}$ accurately. At larger time scales, however, the spectrum becomes
dense. This means that the number of states needed for an accurate
representation starts to increase exponentially in agreement with our
numerical findings shown in Fig.\ref{fig3}(b). The breakdown approximately
occurs when the discarded weight defined by $1-\sum_{i=1}^Z \Lambda_i$, where
$\Lambda_i$ are the $Z$ largest eigenvalues of $\rho_S$, becomes larger than
$10^{-3}$ (see inset of Fig.~\ref{fig3}(c)). The long-time asymptotics of the
autocorrelation function at $T=\infty$ is therefore not accessible 
%
%
within our method.

\begin{figure}[!htp]
\begin{center}
\includegraphics*[width=0.99\columnwidth]{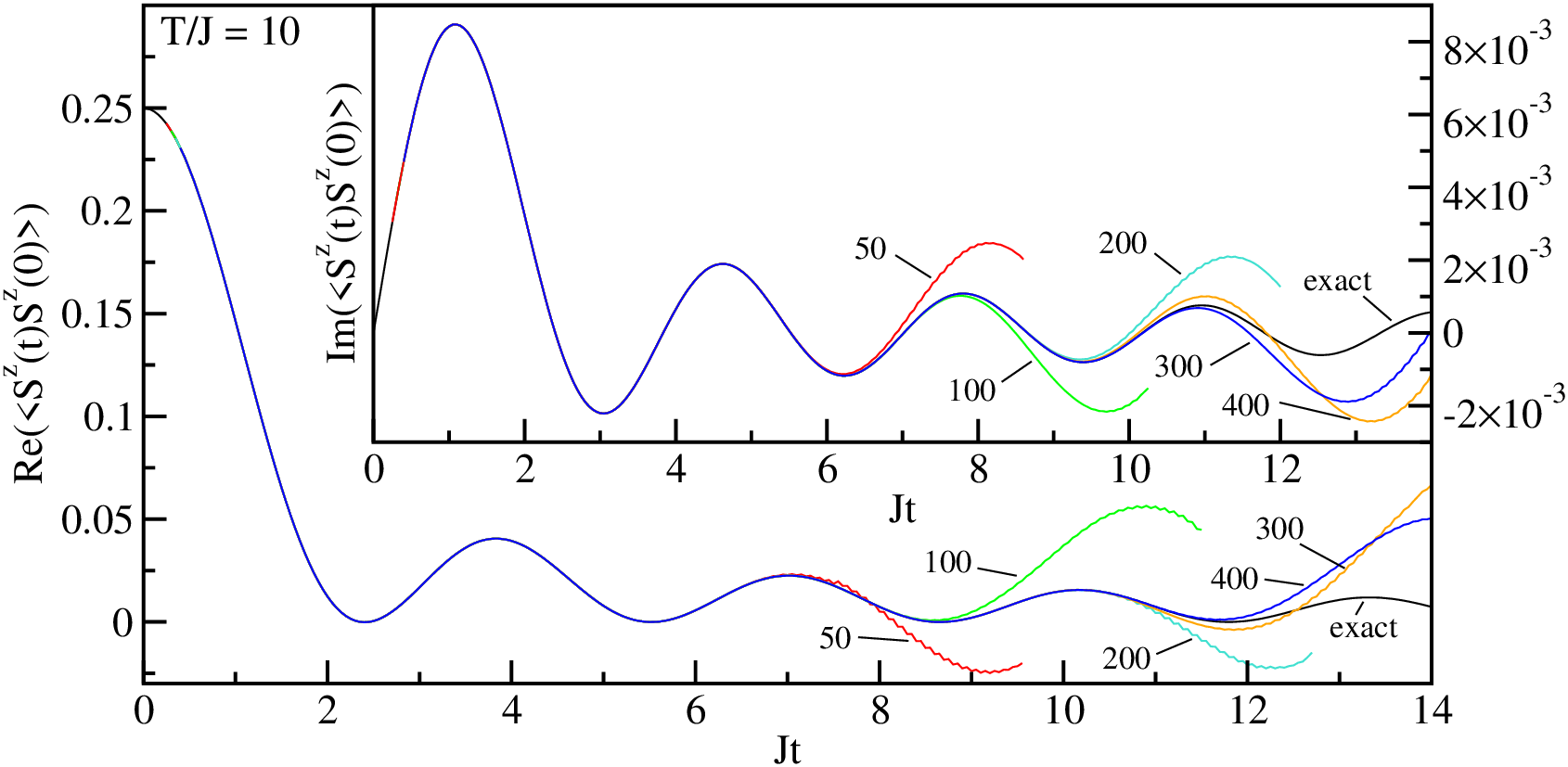}
\includegraphics*[width=0.99\columnwidth]{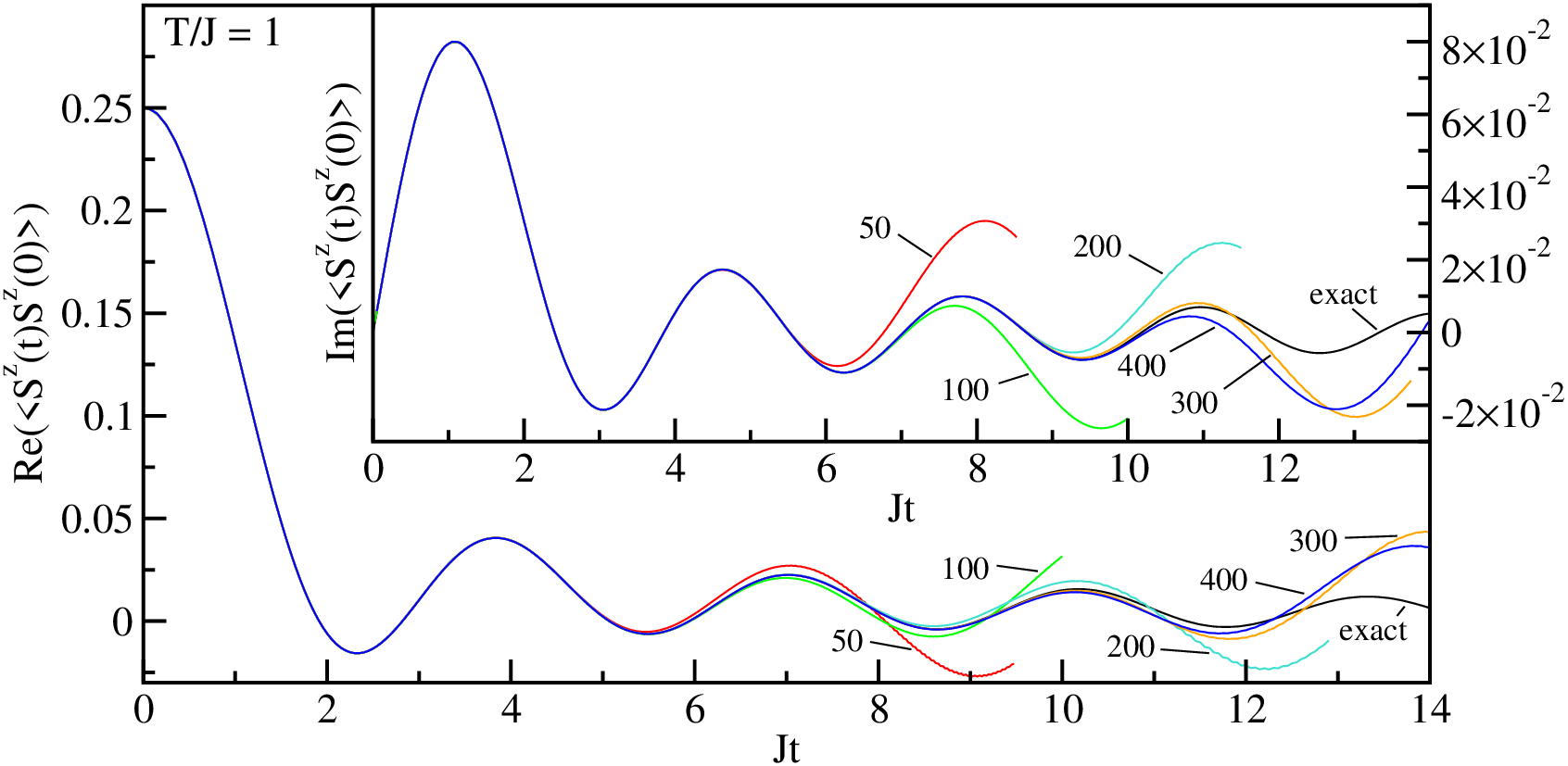}
\includegraphics*[width=0.99\columnwidth]{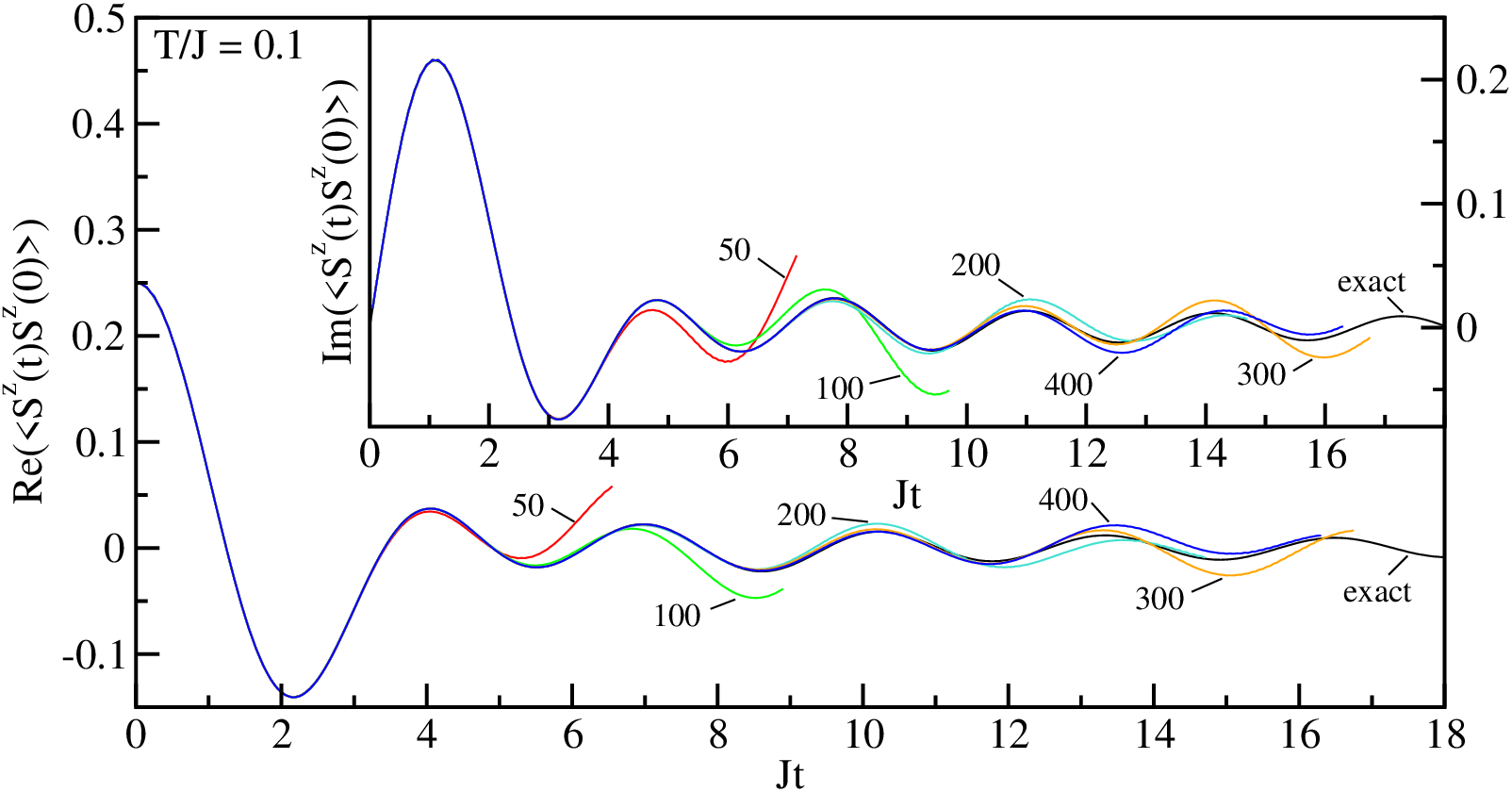}
\end{center}
\caption{$C(t,T)$ for $\Delta=0$ at $T/J=10,1,0.1$. The main
  figures show the real, the insets the imaginary parts.}
\label{fig4}
\end{figure}

Next, we consider finite temperatures $0<T<\infty$. Results for $\Delta=0$ and
$T/J=10,1,0.1$ are shown in Fig.~\ref{fig4}. According to Eq.~(\ref{eq9}),
$C(t,T)$ now acquires also 
%
%
an imaginary part which is shown in the insets of Fig.~\ref{fig4}.
For $T/J=10$ and $T/J=1$ the results look qualitatively similar to the
$T=\infty$ case. As shown in Fig.~\ref{fig3}(b) the number of states needed to
obtain the same accuracy at a given time increases with decreasing
temperature.  This is easy to understand because the Hilbert space for
$\mathcal{T}$ increases when adding additional $\tau$-plaquettes. The
breakdown of the algorithm also looks similar to the $T=\infty$ case and we
again find an exponential increase in $Z_n$ at larger times. As we have chosen
$\epsilon=\delta$, the number of $\tau$-plaquettes $M$ is much smaller than
the number of $v$-plaquettes $N$ for $t\sim 10$ where the calculations with
$Z=300,400$ states fail. The spectrum of the reduced density matrix $\rho_S$
at these time scales will therefore look very similar to the one for
$T=\infty$ shown in Fig.~\ref{fig3}(c). We therefore conclude that it is again
the spectrum of $\rho_S$ which sets the limiting time for our calculations.

For $T/J=0.1$, however, we find a different behavior. Instead of a rapid
breakdown with arbitrarily large deviations for $t>t_c$ we find that the
results for all $Z$ remain relatively close to the exact one over the entire
time scale investigated (the data for $Z=50,100$ in Fig.~\ref{fig4} are
%
depicted only up to times marking the start of deviations). We also see from
Fig.~\ref{fig3}(b) that the functional form of the increase in $Z_n$ is now
different from the cases $T/J=\infty,10,1$. Whereas it becomes exponential in
the latter, it is more or less linear for $T/J=0.1$
%
%
%
from $Z_n=200$ to $Z_n=400$. This can be understood as follows: For a transfer
matrix $\mathcal{T}$ consisting only of $\tau$-plaquettes, the spectrum of
$\rho_S$ is exponentially decaying. This characteristic should be preserved as
long as the number of $v$-plaquettes is not much larger than the number of
$\tau$-plaquettes. A fundamental failure of our approach due to a dense
spectrum of $\rho_S$ should only occur for $N\gg M$. By considerably
increasing $Z$ it should therefore be possible to access much larger time
scales at low temperatures in future large scale numerical studies.

To conclude, we have presented a numerical method to calculate real-time
correlations in one-dimensional quantum systems at finite temperature. The
method is based on a DMRG algorithm applied to transfer matrices. As
essentially new ingredient it involves a second Trotter-Suzuki decomposition
for the time evolution operator. To test our approach we have calculated the
autocorrelation function for the $XXZ$-chain both at infinite and finite
temperatures. For $T=\infty$ we have established that reliable results can be
obtained up to a maximum time scale $t_c$ where the spectrum of the reduced
density matrix $\rho_S$ becomes dense. For low $T$ we have shown that the
algorithm does not show
%
%
a rapid breakdown contrary to the high-$T$ case and have argued that the
fundamental problem of $\rho_S$ becoming dense is less severe. A huge
advantage of our approach compared to other methods is that once the blocks
necessary to construct $\mathcal{T}, \mathcal{T}(O)$ at a given time $t$ are
known, the CF can be evaluated at time $t$ for arbitrary distances $r$ between
the operators $O$ by simple matrix-vector multiplications. As our approach is
working directly in the thermodynamic limit it is possible to obtain highly
accurate results even for large distances as has been demonstrated for the
static case in Ref.~\onlinecite{SirkerKluemperPRB}. This will be exploited in
a forthcoming publication \cite{inprep} for a detailed study of time ($t<t_c$)
and space dependent CFs in the $XXZ$ chain.

\acknowledgments The authors acknowledge valuable discussions with I.~Peschel
and R.~Noack and thank K.~Fabricius for providing full diagonalization data
for comparison. JS acknowledges support by the DFG.  

\end{document}